\documentclass[12pt,aps,prd,tightenlines,nofootinbib,floatfix]{revtex4-2}
\usepackage{latexsym,amsmath,amssymb,amsbsy,graphicx}
\usepackage{multirow}
\usepackage{rotating}
\usepackage{xcolor}

\begin{document}
	
\title{Exclusive semileptonic and nonleptonic $J/\psi$ decays}

\author{V.\,O. \surname{Galkin$^1$}} 
\email{vgalkin@frccsc.ru}
\author{I.\,S. \surname{Sukhanov$^{1,2}$}}
\email{sukhanov.is17@physics.msu.ru}

\affiliation{$^1$
	Federal Research Center ``Computer Science and Control'' Russian Academy of Sciences, Vavilov Street 40, Moscow, 119333 Russia}
\affiliation{$^2$  Faculty of Physics, Lomonosov Moscow State University, Leninskie Gory 1-2,
	Moscow 119991, Russia.}

%\date{\today}

\preprint{}	
	
\begin{abstract}
Exclusive semileptonic and nonleptonic $J/\psi$ decays are investigated in the framework of the relativistic quark model based on the quasipotential approach and quantum chromodynamics. The form factors parameterizing the hadronic matrix element of the weak current are calculated with the complete account of the relativistic effects. These form factors are expressed as the overlap integrals of the meson wave functions and are determined in the whole accessible kinematic range. On this basis the semileptonic decay branching fractions are evaluated for decays involving both electrons and muons. The nonleptonic decays are considered in the factorization approximation in the limit for the number of colors $N_c\to \infty$. The obtain branching fractions are found to be of the order $10^{-9}\sim 10^{-12}$. They are compared with the previous theoretical predictions and  available experimental upper bounds.  

\end{abstract}

\maketitle

\section{INTRODUCTION} \label{sec:intro}
The $J/\psi$ meson is a bound state of the charm quark and charm antiquark  ($c\overline{c}$) whose mass is below the threshold of the $D\overline{D}$ production. As a result, the $J/\psi$ meson has a small total decay width  $\Gamma_{J/\psi}=92.6\pm1.7$~keV. Its
decays are dominated by strong ($J/\psi\to ggg$) and electromagnetic ($J/\psi\to\gamma gg$) interactions, therefore, possible weak decays are significantly suppressed. However, given that the BESIII experiment has already accumulated over 10 billion $J/\psi$ events \cite{Ablikim2022}, and more data are expected to come from the future Super Tau Charm Facility \cite{Achasov2024}, these weak decays could be measured in the near future.

The $J/\psi$ meson can weakly decay into a single charm meson, accompanied by light hadrons or leptons, via the weak decay of one of the charm quarks. The Standard Model (SM) predicts that the inclusive branching fraction of such weak decays  is at best of the order of $10^{-8}$ or even lower \cite{SanchisLozano1994}. Thus, searching for these decays not only tests the SM prediction but also explores new physical theories beyond the SM, since the possible ``new physics" contributions can significantly increase the corresponding branching fractions. Until now, weak decays of the $J/\psi$ meson have not been observed, but experimental constraints on semileptonic \cite{Ablikim2024a,Ablikim2021,Ablikim2006,Ablikim2014}, nonleptonic \cite{Ablikim2024,Ablikim2025a,Ablikim2025b}, and rare \cite{Ablikim2017,Ablikim2025,Ablikim2006} weak $J/\psi$ decays have been significantly improved in recent years. The recent review of experimental data and theoretical predictions is given in Ref.~\cite{Li:2026}.

The weak decays of the heavy quarkonium-like $J/\psi$ meson offer an ideal opportunity for studying nonperturbative QCD effects, because such systems contain two heavy constituents of the same flavor. Moreover, for the weak decay of a vector meson, such as $J/\psi$, the polarization effect may play a role in testing the underlying dynamics and hadron structure.
In the weak decay of the $J/\psi$ one of the heavy charm  quarks ($c$ or $\bar c$) weakly decays, while the other one ($\bar c$ or $c$) acts as a spectator. Therefore, the final state contains a single charmed hadron. The effective Hamiltonian governing the $c$ quark weak decay processes is well known. The main theoretical problem while calculating the semileptonic and nonleptonic decay  rates of the $J/\psi$ consists in evaluation of the hadronic matrix elements of the weak current, which is governed by the nonperturbative QCD dynamics. These transition matrix elements are parameterized by the set of the invariant form factors. 
To calculate the form factors of weak decays the nonperturbative QCD approaches should be used. They include lattice QCD, sum rules and various quark models. It is necessary to calculate these form factors not only in some fixed kinematical point (usually at the point of zero recoil of the final meson, $q^2=0$), but reliably determine their momentum transfer dependence in the whole kinematical range. Form factor values vary greatly depending on the model, which represents an additional test of the employed approaches.

For theoretical calculations of the nonleptonic decays of the $J/\psi$ meson the factorization approach, where the hadronic matrix element is factorized into a product of two matrix elements of single currents,  is usually used. Nonfactorizable effects are included in the effective coefficients, which are assumed to be universal. Their perturbative account is possible only in some specific cases. As a result these coefficients are process dependent, but the changes in their values may not be very significant. For weak decays of heavy mesons, this factorization approach has shown  good results for the color allowed processes. Thus one can expect that this assumption can also hold for the nonleptonic $J/\psi$ decays governed by the color allowed processes. However, for the nonleptonic decay processes of charm mesons, in which color is suppressed, the obtained predictions are orders of magnitude smaller than the experimental values. Therefore, for the description of such processes, we  use the approximation of the infinite number of colors ($N_c\to\infty$) proposed in Ref~\cite{Buras1995}.

The paper is organized as follows. In Section~\ref{sec:rqm} we briefly describe our relativistic quark model. The calculation of the matrix elements of the weak current between meson states with the account of relativistic effects is discussed. In Section~\ref{sec:sd}  this model is applied to the calculation of the form factors parameterizing the hadronic matrix elements for the weak decay of $J/\psi$ to open charm mesons. First,  semileptonic $J/\psi$ decays are discussed. Then in Sec.~\ref{sec:non} the  nonleptonic $J/\psi$ decays are considered in the factorization approximation.   The numerical results for the form factors and branching fractions of the semileptonic and nonleptonic $J/\psi$ decays are presented in Section~\ref{sec:rd}. They are compared with previous theoretical predictions and available experimental upper bounds.  Finally, Section~\ref{sec:con} contains our conclusions.

\section{RELATIVISTIC QUARK MODEL}	\label{sec:rqm}
	
We perform all calculations in the relativistic quark model (RQM) based on the quasipotential approach. First, we need to obtain the initial and final meson wave functions. In RQM the wave function $\Psi_M(\textbf{p})$ of the meson with mass $M$, which is considered to be the quark-antiquark bound state, satisfies the relativistic Schr\"{o}dinger-like quasipotential equation \cite{Ebert2003}	
\begin{equation}\label{1}
\Bigg(\dfrac{b^2(M)}{2\mu_R}-\dfrac{\textbf{p}^2}{2\mu_R}\Bigg)\Psi_M(\textbf{p})=\int \dfrac{d^3q}{2\pi^3}V(\textbf{p},\textbf{q};M)\Psi_M(\textbf{q}).
\end{equation}
Here $\textbf{p}$ is the relative momentum of quarks with the constituent masses $m_{1,2}$, while $b^2(M)$ and $\mu_R$ are the relative momentum squared in the center of mass system on the mass shell and the relativistic reduced mass, respectively, defined by
\begin{eqnarray}
\label{2}
b^2(M) &=& \dfrac{\big[ M^2-(m_1+m_2)^2\big]\big[M^2-(m_1-m_2)^2\big]}{4M^2},\\
\label{3}
\mu_R &=& \dfrac{M^4-(m^2_1-m^2_2)^2}{4M^3}.
\end{eqnarray}
The left hand side of this equation contains the relativistic kinematics, which leads to the complicated dependence of the reduced mass $\mu_R$ and the equation eigenvalue [$b^2(M)/2\mu_R$] on the meson mass $M$. The relativistic dynamics is contained in the kernel of this equation $V(\textbf{p},\textbf{q};M)$, which is the QCD-motivated quark-antiquark potential.  We construct this quasipotential by the off-mass-shell scattering amplitude projected on the positive energy states \cite{Ebert2003}.  It includes the short range one-gluon exchange term and a long range confining potential considered to be a mixture of the scalar and vector linear confining interactions. The long-range vertex of the confining vector interaction contains an additional Pauli term. The resulting quasipotential is given by 	\cite{Ebert2003}
\begin{eqnarray}\label{Ker}
V(\textbf{p},\textbf{q};M)=\bar u_1(\textbf{p)} \bar u_2(-\textbf{p})&\!\!\Biggl\{\!\!&\frac{4}{3}\alpha_s D_{\mu \nu}(\textbf{k})\gamma^\mu_1\gamma^\nu_2\cr
&&  +V^V_{\rm conf}(\textbf{k})\Gamma^\mu_1(\textbf{k})\Gamma_{2;\nu}(\textbf{k})+V^S_{\rm conf}(\textbf{k})\Biggr\} u_1(\textbf{q})u_2(-\textbf{q}),
\end{eqnarray}
where $\alpha_s$ is the QCD coupling constant, $D_{\mu\nu}$ is the gluon propagator in the Coulomb gauge, $\textbf{k}=\textbf{p}-\textbf{q}$, and $\gamma_\mu$ and $u({\bf p})$ are the Dirac matrices and spinors, respectively. The long-range vector vertex consists of both Dirac and Pauli terms
\begin{equation}
\Gamma_\mu(\textbf{k})=\gamma_\mu+\frac{i\kappa}{2m}\sigma_{\mu \nu}k^\nu,
\end{equation}
where $\kappa$ is the long-range anomalous chromomagnetic quark moment. In the nonrelativistic limit confining vector and scalar potentials are given by
\begin{equation}
V^V_{\rm conf}(r)=(1-\varepsilon)(Ar+B), \qquad V^S_{\rm conf}(r)=\varepsilon(Ar+B),
\end{equation}
with the mixing coefficient $\varepsilon$. The emerging quasipotential can be viewed as the relativistic generalization of the nonrelativistic Cornell potential
\begin{equation}
V_{NR}(r)=-\frac{4}{3}\frac{\alpha_s}{r}+Ar+B,
\end{equation}
which contains both spin-independent and spin-dependent relativistic contributions.

All parameters of the model were fixed from the previous consideration of hadron spectroscopy and decays \cite{Ebert2003}. They are the following: the  constituent quark masses are $m_c = 1.55$~GeV, $m_s =0.5$~GeV, $m_{u,d} = 0.33$~GeV; the parameters of the linear potential are $A = 0.18$~GeV$^2$ and $B = -0.30$~GeV; the mixing parameter of the scalar and vector confining potentials $\varepsilon=-1$ and the universal Pauli interaction constant $\kappa=-1$.
In our calculations we take the running QCD coupling constant with infrared freezing \cite{Ebert2009}.

Second, we use these wave functions to study the weak decays of the $J/\psi$ meson into the final charm meson ($D$ or $D_s$). These decays are governed by the weak current $J^W_\mu=\overline{f}\gamma_\mu(1-\gamma_5)c$, where the final quark $f = s, d$. Thus it is necessary to calculate the corresponding hadronic matrix element of this local current operator between the initial $J/\psi$ meson  and the final $D_{(s)}$  meson. This matrix element in the quasipotential approach is expressed \cite{Ebert2003a} through the overlap integral of the initial $J/\psi$ and final $D_{(s)}$  meson wave functions $\Psi_{M\textbf{p}_M}$ ($M=J/\psi$ or $D_{(s)}$), which are projected on the positive energy states and boosted to the moving reference frame with the three-momentum $\textbf{p}_M$  
\begin{equation}\label{Matrix}
\langle{D_{(s)}(p_{D_{(s)}})|J^W_\mu|J/\psi(p_{J/\psi})}\rangle = \int \dfrac{d^3p d^3q}{(2\pi)^6} \overline{\Psi}_{D_{(s)}\textbf{p}_{D_{(s)}}}(\textbf{p})\Gamma_\mu(\textbf{p},\textbf{q})\Psi_{J/\psi\textbf{p}_{J/\psi}}(\textbf{q}).
\end{equation}
Here  the vertex function $\Gamma=\Gamma^{(1)}+\Gamma^{(2)}$ contains two contributions. The first one $\Gamma^{(1)}$ is the leading-order vertex function corresponding to the impulse approximation, while the second one $\Gamma^{(2)}$ is the consequence of the projection on the positive energy states in the quasipotential approach. It takes into account interaction of the active quarks  with the spectator antiquark  and includes the negative-energy part of the active quark propagator. The explicit form of these vertex functions can be found in Ref.~\cite{Ebert2003a}.

The meson wave functions, obtained by solving the quasipotential equation (\ref{1}), are in the meson rest frame. However, the relativistic meson wave functions in the matrix element (\ref{Matrix}) should be taken in the reference frame moving with the meson momentum. It is reasonable to use the rest frame of the decaying hadron, the $J/\psi$ in our case. Then its momentum ${\bf p}_{J/\psi} = 0$ and the final meson $D_{(s)}$ is moving with the recoil momentum ${\bf\Delta} ={\bf p}_{D_{(s)}}$ and its wave function should be boosted to the moving reference frame. The wave function of the moving meson $\Psi_{D_{(s)}\,{\bf\Delta} }$ is connected with the wave function in the rest frame $\Psi_{D_{(s)}\,{\bf 0}}$ by the transformation \cite{Ebert2003a}
\begin{equation}
\label{wig}
\Psi_{D_{(s)}\,{\bf\Delta}}({\bf
	p})=D_q^{1/2}(R_{L_{\bf\Delta}}^W)D_{\bar q}^{1/2}(R_{L_{
		\bf\Delta}}^W)\Psi_{D_{(s)}\,{\bf 0}}({\bf p}),
\end{equation}
where $R^W$ is the Wigner rotation, $L_{\bf\Delta}$ is the Lorentz boost
from the meson rest frame to a moving one, and  $D^{1/2}_q(R)$ is 
the rotation matrix  in spinor representation.

\section{Semileptonic decays}	\label{sec:sd}
	
  The effective Hamiltonian, $H_{\rm eff}(c\to ql\nu_l)$ $(q=s,d)$, describing the semileptonic $J/\psi$ decays to  $D_{(s)}$ mesons can be expressed  in the Standard Model as follows
\begin{equation}
H_{\rm eff}(c\to ql\nu_l)=\dfrac{G_F}{\sqrt{2}}V_{cq}[ \overline{q}\gamma^\mu(1-\gamma_5)c][ \overline{\nu}_l\gamma_\mu(1-\gamma_5)l],
\end{equation}
where  $V_{cq}$ is the corresponding Cabibbo-Kobayashi-Maskawa (CKM) matrix element and $G_F$ is the Fermi constant.
Its matrix element $\mathcal{M}$ between the initial ($J/\psi$) and final ($D_{(s)}l\nu_l$) states factorizes in the product of the leptonic current $L_\mu=\overline{\nu}_l\gamma_\mu(1-\gamma_5)l$ and the matrix element of the hadronic current $H^\mu=\langle{D_{(s)}}| \overline{q}\gamma_\mu(1-\gamma_5)c|{J/\psi}\rangle$  	
	\begin{equation} \label{me}
	\mathcal{M}(J/\psi\to D_{(s)}l\nu_l)=\dfrac{G_F}{\sqrt{2}}V_{cq}H^\mu L_\mu.
	\end{equation}
 The leptonic part is calculated using the lepton spinors and has a simple structure. The hadronic part is significantly more complicated since its calculation requires nonperturbative treatment within QCD. 	

	It is convenient to parameterize the hadronic matrix element of the weak current $J^W$ between meson states for the $J/\psi$ transition to the $D_{(s)}$ meson by the following set of the invariant form factors \cite{Galkin2025}:
	\begin{eqnarray}\label{ff}
	\langle D_{(s)}(p_{D_{(s)}})|\overline{q}\gamma^\mu c|J/\psi(p_{J/\psi})\rangle & =&
		\displaystyle{\frac{2iV(q^2)}{M_{D_{(s)}}+M_{J/\psi}}}{\epsilon}^{\mu \nu \rho \sigma}{\epsilon}_\nu p_{J/\psi\rho}p_{D_{(s)}\sigma},\\
\!\!\!\!\!		\nonumber\langle D_{(s)}(p_{D_{(s)}})|\overline{q}\gamma^\mu \gamma_5c|J/\psi(p_{J/\psi})\rangle &=& 2M_{D_{(s)}}A_0(q^2)\displaystyle\frac{\epsilon \cdot q}{q^2}q^\mu%\\&&\nonumber
		+(M_{D_{(s)}}+M_{J/\psi})A_1(q^2)\Big(\epsilon^{\mu}-\displaystyle\frac{\epsilon \cdot q}{q^2}q^\mu\Big)\\&&\!\!\!\!\! -A_2(q^2)\displaystyle\frac{\epsilon \cdot q}{M_{D_{(s)}}+M_{J/\psi}}\left[p^\mu_{D_{(s)}}+p^\mu_{J/\psi}-\dfrac{M^2_{J/\psi}-M^2_{D_{(s)}}}{q^2}q^\mu\right].\nonumber\qquad
	\end{eqnarray}
Here the momentum transferred to the lepton pair $q=p_{J/\psi}-p_{D_{(s)}}$. The following relations among form factors should be satisfied to avoid singularities at the maximum recoil~($q^2=0$) 
	\[A_0(0)=\frac{M_{D_{(s)}}+M_{J/\psi}}{2M_{D_{(s)}}}A_1(0)-\frac{M_{J/\psi}-M_{D_{(s)}}}{2M_{D_{(s)}}}A_2(0).\]
We first consider the twofold differential decay distribution in terms of $q^2$ and the polar angle $\theta$. The polar angle $\theta$ is defined in the rest frame of the $W^*$ as the angle between the momentum of the final charged lepton and the direction opposite to the daughter meson's momentum. The differential distribution of the $J/\psi$ semileptonic decay to the final $D_{(s)}$ meson is given by
\begin{eqnarray}\label{Gamma}
\nonumber \dfrac{d\Gamma_{J/\psi\to D_{(s)}l\nu_l}}{dq^2d\cos\theta}&=&\dfrac{\lambda^{1/2}(M_{J/\psi}^2,M_{D_{(s)}}^2,q^2)}{(2\pi)^3\ 24M_{J/\psi}^3}(1-2\delta_l)\bar{\sum_{\rm pol}}|\mathcal{M}(J/\psi\to D_{(s)}l\nu_l)|^2\\
&=&\dfrac{1}{3}\dfrac{G_F^2}{(2\pi)^3}\dfrac{\lambda^{1/2}(M_{J/\psi}^2,M_{D_{(s)}}^2,q^2)}{128M_{J/\psi}^3}|V_{cq}|^2(1-2\delta_l)H^{\mu\nu}L_{\mu\nu},
\end{eqnarray}
where  the square of the weak decay matrix element $\mathcal{M}(J/\psi\to D_{(s)}l\nu_l)$ (\ref{me}) is summed over the final state  and averaged over the initial state polarizations. The average over the spin states of the initial vector $J/\psi$ meson gives the factor $1/3$. Here $\lambda(x,y,z)= x^2+y^2+z^2-2(xy+xz+yz)$ is the K\"all\'en function, $\delta_l=\frac{m_l^2}{2q^2}$ and $m_l$ is the lepton mass.
Integrating over $\cos\theta$ and employing the helicity amplitudes of the hadronic and leptonic tensors one obtains the differential decay distribution over $q^2$
\begin{equation}
\dfrac{d\Gamma_{J/\psi\to D_{(s)}l\nu_l}}{dq^2}=\dfrac{1}{3}\dfrac{G^2_F}{(2\pi)^3}|V_{cq}|^2\dfrac{\lambda^{1/2}(M_{J/\psi}^2,M_{D_{(s)}}^2,q^2)q^2(1-2\delta_l)^2}{24M_{J/\psi}^3}H_{\rm tot},
\end{equation}
where the total helicity structure $H_{\rm tot}$ is defined by
\begin{equation}
H_{\rm tot}=(1+\delta_l)(|H_+(q^2)|^2+|H_-(q^2)|^2+|H_0(q^2)|^2)+3\delta_l|H_t(q^2)|^2.
\end{equation}
The subscripts $\pm,0,t$ denote transverse, longitudinal, and time helicity components, respectively. They are expressed through the form factors (\ref{ff})  as follows \cite{Faustov2019}
\begin{eqnarray}\label{hs}
H_\pm(q^2)&=&\dfrac{\lambda^{1/2}(M_{J/\psi}^2,M_{D_{(s)}}^2,q^2)}{M_{D_{(s)}}+M_{J/\psi}}\left[V(q^2)\mp \dfrac{(M_{D_{(s)}}+M_{J/\psi})^2}{\lambda^{1/2}(M_{J/\psi}^2,M_{D_{(s)}}^2,q^2)}A_1(q^2)\right],\cr
H_0(q^2)&=&\dfrac{1}{2M_{D_{(s)}}\sqrt{q^2}}\Bigg[(M_{D_{(s)}}+M_{J/\psi})(M_{J/\psi}^2-M^2_{D_{(s)}}-q^2)A_1(q^2)\cr
&&\phantom{\dfrac{1}{2M_{J/\psi}\sqrt{q^2}}\Bigg[}-\dfrac{\lambda(M_{J/\psi}^2,M_{D_{(s)}}^2,q^2)}{M_{D_{(s)}}+M_{J/\psi}}A_2(q^2)\Bigg],\cr
H_t(q^2)&=&\dfrac{\lambda^{1/2}(M_{J/\psi}^2,M_{D_{(s)}}^2,q^2)}{\sqrt{q^2}}A_0(q^2).
\end{eqnarray}

\section{Nonleptonic decays} \label{sec:non}
In the Standard Model, the effective Hamiltonian of the weak nonleptonic decay of the meson with the charm quark is given by \cite{Yu2023}:
\begin{equation}
H_{\rm eff}=\dfrac{G_F}{\sqrt{2}}V_{cq_1}^*V_{uq_2}\Big[C_1(\mu)Q_1+C_2(\mu)Q_2\Big],
\end{equation}
where $q_{1,2}$ are $s$ or $d$ quarks, $C_1$ and $C_2$ are Wilson coefficients. The operators $Q_i$ ($i=1,2$) are defined as
\begin{eqnarray}
Q_1=(\overline{q}_{1\alpha}c_\alpha)_{V-A}(\overline{u}_\beta q_{2\beta})_{V-A},\\
Q_2=(\overline{q}_{1\alpha}c_\beta)_{V-A}(\overline{u}_\beta q_{2\alpha})_{V-A},
\end{eqnarray}
where $\alpha$ and $\beta$ represent the color indices. The penguin contributions are highly suppressed for charm decays, thus we neglect them. Employing the Fierz transformation
\begin{equation}
\delta_{\alpha\beta}\delta_{\gamma\delta}=\frac{1}{N_c}\delta_{\gamma\beta}\delta_{\alpha\delta}+2T^a_{\alpha\delta}T^a_{\gamma\beta},
\end{equation}
where $N_c$ is the number of colors and $T^a$ are $SU(3)$ color generators,  we can obtain the effective Hamiltonians
corresponding to the color-favored and color-suppressed decay processes. The additional terms with the nonfactorizable color-octet current operators $\overline{Q}_1=(\overline{q}_{1\alpha}T^aq_{2\alpha})_{V-A}(\overline{u}_\beta T^ac_\beta)_{V-A}$ and
$\overline{Q}_2=(\overline{q}_{1\alpha}T^ac_\alpha)_{V-A}(\overline{u}_\beta T^aq_{2\beta})_{V-A}$ emerge, which are neglected in the factorization approach. 

As the result, we can write the effective Hamiltonian corresponding to the color-favored decay process
\begin{equation}
H_{\rm cf}=\dfrac{G_F}{\sqrt{2}}V_{cq_1}^*V_{uq_2}a_1(\overline{q}_{1}c)_{V-A}(\overline{u} q_{2})_{V-A},
\end{equation}
and for the color-suppressed case
\begin{equation}
H_{\rm cs}=\dfrac{G_F}{\sqrt{2}}V_{cq_1}^*V_{uq_2}a_2(\overline{q}_{1}q_2)_{V-A}(\overline{u}c)_{V-A},
\end{equation}
where 
\begin{equation}\label{a}
a_1(\mu)=C_1(\mu)+\dfrac{1}{N_c}C_2(\mu),\qquad a_2(\mu)=C_2(\mu)+\dfrac{1}{N_c}C_1(\mu).
\end{equation}

The simplicity of the factorization approximation is quite attractive, however, it is well known that nonfactorizable contributions must be present in the hadronic matrix elements of the current-current operators $Q_1$ and $Q_2$ in order to compensate for the dependence of the coefficients $C_i(\mu)$ or $a_i(\mu)$ on the scale $\mu$, since the physical amplitudes must be independent of an arbitrary renormalization scale $\mu$ \cite{Buras1995}. In the factorization approach $\langle Q_i\rangle_F$ are the products of matrix elements of conserved currents, which are independent of $\mu$. As a result there is no compensation for the dependence on $\mu$ of the Wilson coefficients. Therefore, the factorization approach can at best be correct for a single value of $\mu$, the so-called factorization scale $\mu_F$. Note that the approach itself does not fix a value for the factorization scale $\mu_F$. Nevertheless, it is usually assumed that $\mu_F$ is of the order of the mass of the decaying quark, the $c$ quark in the considered case  [$\mu_F = O(m_c)$]. As it is pointed out in Ref.~\cite{Buras1995}, at next to leading order in the renormalization group improved perturbation theory the coefficients $C_i(\mu)$ acquire dependence on the renormalization scheme for operators. Again only the presence of non-factorizable contributions in $\langle \overline{Q}_i\rangle$ can remove this scheme dependence in the physical amplitudes. However, $\langle \overline{Q}_i\rangle_F$ are independent of the renormalization scheme, and the factorization approach cannot unequally identify the preferred renormalization scheme.

It is shown in Ref.~\cite{Buras1995} that in the leading order of perturbation theory the value of the coefficient $a_2$ is generally substantially smaller than its phenomenological value due to the strong cancellation between $C_2$ and $C_1/3$. One finds typically $a_2 = O(0.1)$ and consequently branching ratios for the nonleptonic decays substantially smaller (almost by an order of magnitude) than the experimental branching ratios.
Therefore  in factorization calculations the limit $N_c\to\infty$ is used, thus discarding ``$1/N_c$” terms in Eq.~(\ref{a}). Using the values of Wilson coefficients \cite{Boer2016} at the scale $\mu=m_c$, we get the following values of  $a_1=1.27$ and $a_2=-0.52$.

In the factorization approach, the hadronic matrix element $$X_{J/\psi\to D_{(s)}M}=\langle D_{(s)}|(\bar q c)_{V-A}|J/\psi\rangle\langle M|(\bar u q)_{V-A}|0\rangle$$ can be expressed by the product of the decay constant and the invariant form factors parameterizing the matrix element $\langle D_{(s)}|(\bar q c)_{V-A}|J/\psi\rangle$. The decay constant of the ground state meson is defined through the matrix element of the weak current between the vacuum and a pseudoscalar ($P$) or a vector ($V$) meson
\begin{eqnarray}
\langle P(p_\mu)|(\overline{q}_1^\prime q_2^\prime)_{V-A}|0\rangle&=&-if_Pp_\mu,\\
\langle V|(\overline{q}_1^\prime q_2^\prime)_{V-A}|0\rangle&=&if_VM_V\epsilon^*_\mu,
\end{eqnarray}
where $M_V$ and $\epsilon_\mu$ are the mass and polarization vector of the vector meson; $f_P$ and $f_V $ are the decay constants of the pseudoscalar
and vector meson, respectively; in our calculations we use the following values of the decay constants \cite{Yu2023}: $f_\pi=130.2$~MeV, $f_K=155.6$~MeV,$f_{K^*}=217$~MeV,  $f_\rho=205$~MeV.

For $J/\psi\to D_{(s)}P$ transitions the matrix element $X_{J/\psi\to D_{(s)}P}$ is expressed as \cite{Yu2023}
\begin{equation}\label{x1}
X_{J/\psi\to D_{(s)}P}=i M_Pf_PH_t(M_P^2).
\end{equation}

For $J/\psi\to D_{(s)}V$ transitions the modulus of the matrix element $X_{J/\psi\to D_{(s)}V}$ squared is given by \cite{Yu2023}
\begin{equation}\label{x2}
|X_{J/\psi\to D_{(s)}V}|^2=f_V^2M_V^2(|H_+(M_V^2)|^2+|H_-(M_V^2)|^2+|H_0(M_V^2)|^2),
\end{equation}
where the helicity components are defined in Eq.~(\ref{hs}) in terms of the decay form factors.

The resulting decay amplitudes of the nonleptonic $J/\psi$ decays can be written as follows
\begin{eqnarray}
A_{J/\psi\to D_{(s)}^-\pi^+}&=&-\dfrac{G_F}{\sqrt{2}}V^*_{cq}V_{ud}a_1X_{J/\psi\to D_{(s)}^-\pi^+},\\
A_{J/\psi\to D^0\pi^0}&=&-\dfrac{G_F}{\sqrt{2}}V^*_{cd}V_{ud}a_2X_{J/\psi\to D^0\pi^0},\\
A_{J/\psi\to \overline{D}^0\overline{K}^{0(*)}}&=&-\dfrac{G_F}{\sqrt{2}}V^*_{cs}V_{ud}a_2X_{J/\psi\to \overline{D}^0\overline{K}^{0(*)}},\\
A_{J/\psi\to D_{(s)}^-K^{+(*)}}&=&-\dfrac{G_F}{\sqrt{2}}V^*_{cq}V_{us}a_1X_{J/\psi\to D_{(s)}^-K^{+(*)}},\\
A_{J/\psi\to D_{(s)}^-\rho^+}&=&-\dfrac{G_F}{\sqrt{2}}V^*_{cq}V_{ud}a_1X_{J/\psi\to D_{(s)}^-\rho^+}.\\
A_{J/\psi\to D^0\rho^0}&=&-\dfrac{G_F}{\sqrt{2}}V^*_{cd}V_{ud}a_2X_{J/\psi\to D^0\rho^0}.
\end{eqnarray}
Then the decay rates are given by
\begin{equation}
\Gamma_{J/\psi\to D_{(s)}P(V)}=\dfrac{1}{3}\dfrac{\lambda^{1/2}(M_{J/\psi}^2,M_{D_{(s)}}^2,M^2_{P(V)})}{16\pi M_{J/\psi}^3}|A_{J/\psi\to D_{(s)}P(V)}|^2,
\end{equation}
here the factor $1/3$ again arises from averaging over the $J/\psi$ spin.

\section{Results and discussion}\label{sec:rd} 
\subsection{Decay form factors}\label{sec:ff}

For the evaluation of the decay rates of weak decays of the $J/\psi$ meson we first need to determine the form factors (\ref{ff}), which parameterize the hadronic matrix elements. We perform calculations in the framework of the quasipotential approach and relativistic quark model briefly discussed in Sec.~\ref{sec:rqm}. On this basis we take into account all relativistic effects including the relativistic contributions of the intermediate negative-energy states and relativistic transformations of the meson wave functions. The resulting expressions for the decay form factors have the form of the overlap integrals of initial and final meson wave functions. They are rather cumbersome and are given in Ref.~\cite{Ebert2003a}. For the numerical evaluation of the decay form factors we use the meson wave functions obtained in calculating their mass spectra \cite{Ebert2009,Ebert2010}. This is a significant advantage of our approach since in most of the previous model calculations some phenomenological wave functions (such as Gaussian) were used. Moreover, our relativistic approach allows us to explicitly determine the form factor dependence on the transferred momentum squared $q^2$ in the whole accessible kinematical range without additional approximations and extrapolations.
We find that the numerical results for these decay form factors can be approximated with high accuracy by the following expressions:
	\begin{equation}
	\label{Ap1}
	F(q^2)=\dfrac{F(0)}{\Big(1+\sigma_1\dfrac{q^2}{M_{D^*}^2}+\sigma_2\dfrac{q^4}{M_{D^*}^4}+\sigma_3\dfrac{q^6}{M_{D^*}^6}\Big)},
	\end{equation}
	where $\sigma_{1,2,3}$ are dimensionless fitted parameters and the mass of the vector $D^*$ meson $M_{D^*} = 2.010$~GeV was used for the normalization.

\begin{table}%[tb]
	\caption{Form factors of the weak $J/\psi$ meson transitions into $D$ mesons.}
	\begin{ruledtabular}
		\begin{tabular}{cccccc}
			%\hline \hline
			\text{Form factors}&$F(0)$&$F(q^2_{\rm max})$&$\sigma_1$&$\sigma_2$&$\sigma_3$\\
			\hline
			$V$&0.479&2.180&$-2.706$&3.904&$-6.061$\\
			$A_0$&0.289&0.914&$-2.209$&4.228&$-8.644$\\
			$A_1$&0.380&0.711&$-1.569$&1.730&$-2.342$\\
			$A_2$&0.656&1.445&$-1.901$&3.220&$-5.476$\\
%\hline \hline
		\end{tabular}\label{FF1}
	\end{ruledtabular}
\end{table}

\begin{table}%[tb]
	\caption{Form factors of the weak $J/\psi$ meson transitions into $D_s$ mesons.}\label{FF2}
	\begin{ruledtabular}
		\begin{tabular}{cccccc}
			%\hline \hline
			\text{Form factors}&$F(0)$&$F(q^2_{\rm max})$&$\sigma_1$&$\sigma_2$&$\sigma_3$\\
			\hline
			$V$&0.608&1.829&$-2.573$&2.724&$-4.060$\\
			$A_0$&0.312&1.088&$-2.534$&3.549&$-8.533$\\
			$A_1$&0.428&0.723&$-1.527$&1.736&$-3.151$\\
			$A_2$&0.835&1.554&$-1.757$&$-3.000$&$-6.604$\\
%\hline \hline
		\end{tabular}
	\end{ruledtabular}
\end{table}

\begin{figure}
	\begin{minipage}[h]{0.48\linewidth}
		\center{\includegraphics[width=1\linewidth]{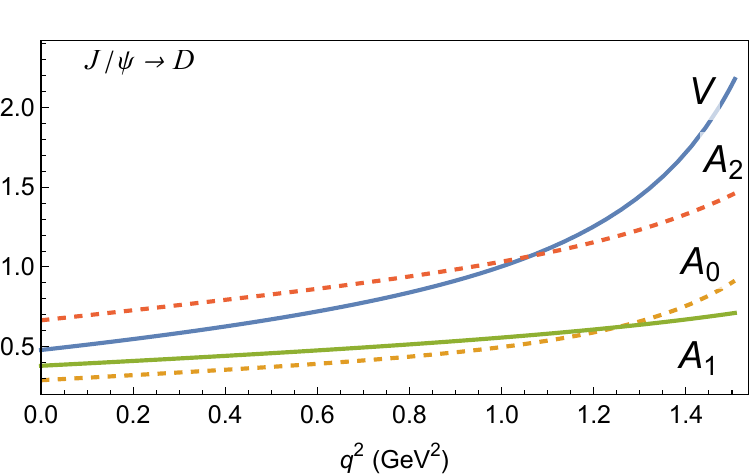}} \\
	\end{minipage}
	\hfill
	\begin{minipage}[h]{0.48\linewidth}
		\center{\includegraphics[width=1\linewidth]{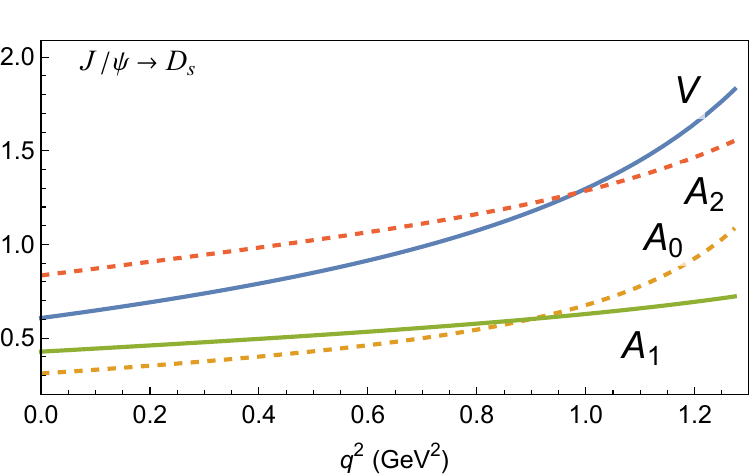}} \\
	\end{minipage}
	\caption{Form factors of the weak transitions of $J/\psi$ mesons into $D_{(s)}$ mesons}
	\label{GrFF}
\end{figure}
\begin{table}%[h!]
	
	\caption{Comparison of the calculated form factors of the $J/\psi\to D$ decays with other theoretical predictions}\label{FFc1}
	\begin{ruledtabular}
		\begin{tabular}{ccccc}
		%\hline\hline
		&$V(0)$&$A_0(0)$&$A_1(0)$&$A_2(0)$\\
		\hline
		\text{Our}&$0.479\pm0.019$&$0.289\pm0.012$&$0.380\pm0.015$&$0.656\pm0.026$\\
		\text{LQCD}\cite{Meng2024}&$1.19\pm0.06$&$0.21\pm0.04$&$0.46\pm0.01$&$1.23\pm0.09$\\
		\text{CCQM}\cite{Ivanov2015}&1.26&0.45&0.44&0.41\\
		\text{BS}\cite{Wang2017}&1.13&0.71&0.48&$-0.23$\\
		\text{QCDSR}\cite{Wang2008a}&0.81&0.45&0.27&$-0.34$\\
		\text{CLFQM}\cite{Shen2008}&$1.6\pm0.1$&$1.13\pm0.02$&$0.68\pm0.01$&$-0.18^{+0.06}_{-0.08}$\\
		\text{BSW}\cite{Dhir2013}&1.17&0.40&0.44&0.23\\
		\text{QCDF}\cite{Sun2016}&$1.76\pm0.03$&$1.41\pm0.02$&$0.72\pm0.01$&$-1.38\pm0.02$\\
		%\hline\hline
		\end{tabular}
	\end{ruledtabular}
\end{table}		

\begin{table}%[h!]
	
	\caption{Comparison of the calculated form factors of the $J/\psi\to D_s$ decays with other theoretical predictions}\label{FFc2}
	\begin{ruledtabular}
		\begin{tabular}{ccccc}
		%\hline\hline
		&$V(0)$&$A_0(0)$&$A_1(0)$&$A_2(0)$\\
		\hline
		\text{Our}&$0.608\pm0.024$&$0.312\pm0.013$&$0.428\pm0.017$&$0.835\pm0.033$\\
		\text{LQCD}\cite{Meng2024}&$1.40\pm0.02$&$0.17\pm0.2$&$0.52\pm0.01$&$1.71\pm0.05$\\
		\text{CCQM}\cite{Ivanov2015}&1.43&0.58&0.56&0.50\\
		\text{BS}\cite{Wang2017}&1.31&0.87&0.60&$-0.34$\\
		\text{QCDSR}\cite{Wang2008a}&1.07&0.58&0.38&$-0.35$\\
		\text{CLFQM}\cite{Shen2008}&$1.8\pm0.0$&$1.07\pm0.02$&$0.68\pm0.01$&$-0.13\pm0.04$\\
		\text{BSW}\cite{Dhir2013}&1.26&0.47&0.55&0.14\\
		\text{QCDF}\cite{Sun2016}&$1.55\pm0.04$&$1.42\pm0.02$&$0.81\pm0.01$&$-1.30\pm0.02$\\
		%\hline\hline
		\end{tabular}
	\end{ruledtabular}
\end{table}
We give the values of form factors $F(0), F(q_{\rm max}^2)$ and parameters $\sigma_{1,2,3}$  fitted to numerically calculated form factors in the whole $q^2$ range in Tables \ref{FF1},~\ref{FF2}. { The uncertainties of the calculated form factors originate form the meson masses, meson wave functions, model parameters, and fitting numerically calculated form factors by Eq. (\ref{Ap1}). They do not exceed 5\%. This estimate is consistent with the comparison of our predictions for the $D$-meson semileptonic decays \cite{Faustov2019} with recent experimental data.} These form factors are plotted in Fig.~\ref{GrFF}.

In Tables \ref{FFc1},~\ref{FFc2} we compare our form factors at zero momentum transfer $q^2=0$ with previous predictions obtained using the lattice QCD (LQCD) \cite{Meng2024}, the covariant constituent quark model (CCQM) \cite{Ivanov2015}, the Bethe-Salpeter equation (BS) \cite{Wang2017}, the QCD sum rules (QCDSR) \cite{Wang2008a}, the covariant light front quark model (CLFQM) \cite{Shen2008}, the Bauer-Stech-Wirbel model (BSW) \cite{Dhir2013} and QCD factorization (QCDF) with the nonrelativistic wave functions with
isotropic harmonic oscillator potential \cite{Sun2016}. We find that the presented values of the form factors significantly differ from each other.  Our predictions for the form factors $V(0)$ are more than a factor of two lower than other values. Our form factors $A_0(0)$ and $A_1(0)$ have  intermediate values. The main differences are in the values of the form factor $A_2(0)$, even the  sign differs. It is positive in our model, LQCD, CCQM and BSW,  while it is negative in BS, QCDSR, CLFQM and QCDF.   The momentum transfer squared $q^2$ behavior of the form factors is also different. Our form factors usually grow more rapidly. Note that the CCQM, CLFQM, QCDF and BSW models use the simple Gaussian meson wave functions, while we employ the wave functions obtained from the solution of the quasipotential equation.

\subsection{Branching fractions}\label{brf}
\begin{figure}
	\begin{minipage}[h]{0.48\linewidth}
		\center{\includegraphics[width=1\linewidth]{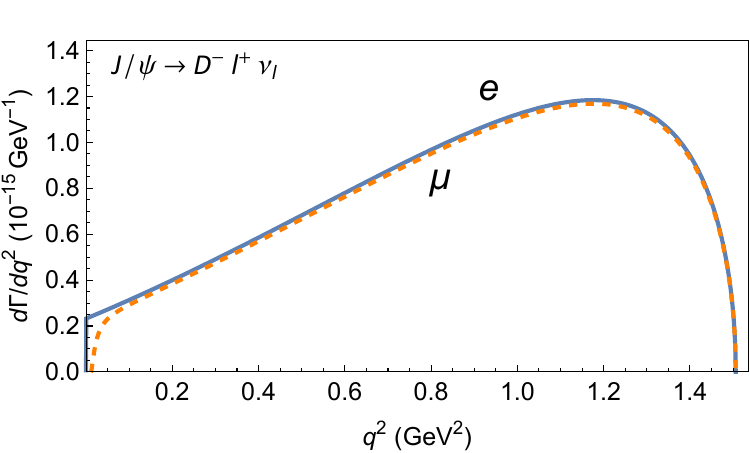}} \\
	\end{minipage}
	\hfill
	\begin{minipage}[h]{0.48\linewidth}
		\center{\includegraphics[width=1\linewidth]{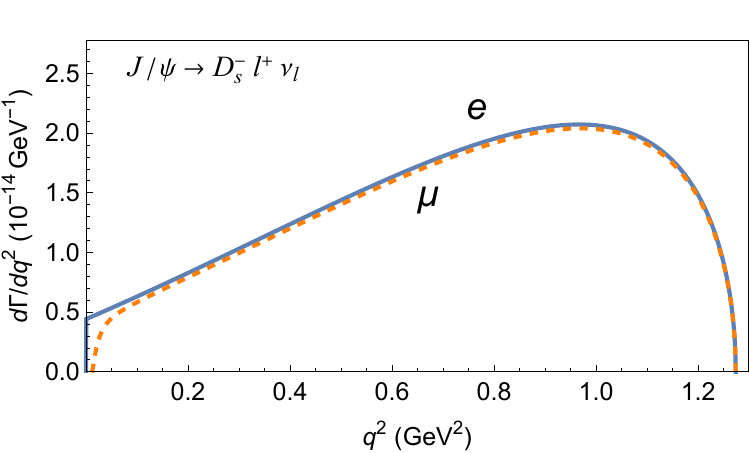}} \\
	\end{minipage}
	\caption{Decay widths of the semileptonic decays of $J/\psi$ mesons into $D_{(s)}$ mesons}
	\label{br}
\end{figure}

We substitute the form factors calculated in Sec.~\ref{sec:ff} in the expressions for the helicity amplitudes (\ref{hs}). Then using helicity components of the hadronic tensor and expression (\ref{Gamma}) for the differential decay rate, we  evaluate the branching fractions of the semileptonic and nonleptonic $J/\psi$ decays. The obtained predictions  are given in Tables~\ref{Br1}--\ref{Br2}. As it is expected, the largest branching fractions (of order of $10^{-10}$) are obtained for the CKM favored $c\to s$ transitions, while for $c\to d$ transitions they are about an order of magnitude smaller. In Fig.~\ref{br} we plot the $J/\psi$ semileptonic decay widths both for electrons (solid line) and muons (dashed line). As we see the main difference between plots occurs for the small values of the transferred momentum squared  $q^2$.

In Table~\ref{Br1} we compare our results for the $J/\psi$ semileptonic decay branching fractions  with the available experimental data \cite{Navas2024} and previous theoretical predictions \cite{Meng2024,Tran2025,Wang2017,Wang2008,Wang2008a,Dhir2013,Shen2008}.
For $J/\psi\to D^-e^+\nu_e$ and $J/\psi\to D^-\mu^+\nu_\mu$ our results are in a good agreement with LQCD and CCQM theoretical predictions of Refs.~\cite{Meng2024,Tran2025} and agree within 2$\sigma$ with BS and QCDSR in Refs.~\cite{Wang2017,Wang2008a}. For the CKM favored $J/\psi\to D_s^-e^+\nu_e$ and $J/\psi\to D_s^-\mu^+\nu_\mu$ decays our results are in good agreement with LQCD and QCDSR theoretical predictions of Refs.~\cite{Meng2024,Wang2008a} and agree within 2$\sigma$ 
with CCQM \cite{Tran2025}. Note that the CLFQM predictions \cite{Shen2008} are more than a factor of two larger than all other results. This is not surprising since CLFQM predicts the largest values of form factors $V(0)$, $A_0(0)$, $A_1(0)$ (see Tables~\ref{FFc1} and \ref{FFc2}).

\begin{sidewaystable}%[h!]
	
	\caption{Branching fraction of the semileptonic $J/\psi\to D_{(s)} \ell\nu_\ell$ decays. }\label{Br1}
	\begin{ruledtabular}
		\begin{tabular}{ccccccccc}
	&\multicolumn{7}{c}{Theory ($\times 10^{-11}$)}&Experiment\\
	\cline{2-8} 
		\text{Decay}&\text{Our}&\text{LQCD} \cite{Meng2024}&\text{CCQM} \cite{Tran2025}&\text{BS} \cite{Wang2017}&\text{QCDSR} \cite{Wang2008a}&\text{BSW} \cite{Dhir2013}&\text{CLFQM} \cite{Shen2008}&\text{PDG} \cite{Navas2024}\\
		\hline
		$J/\psi\to D^-e^+\nu_e$&$1.31\pm0.14$&$1.21\pm0.11$&$1.71\pm0.26$&$2.03^{+0.29}_{-0.25}$&$0.73^{+0.43}_{-0.22}$&$2.3$&$5.1\sim5.7$&$<3.6\times 10^{-8}$\\
		$J/\psi\to D^-\mu^+\nu_\mu$&$1.28\pm0.13$&$1.18\pm0.11$&$1.66\pm0.25$&$1.98^{+0.28}_{-0.24}$&$0.71^{+0.42}_{-0.22}$&$2.2$&$4.7\sim5.5$&$<2.8\times 10^{-7}$\\
		$J/\psi\to D_s^-e^+\nu_e$&$19.9\pm2.1$&$19.0\pm0.8$&$33.0\pm5.0$&$36.7^{+5.2}_{-4.4}$&$18^{+7}_{-5}$&$38.9$&$53\sim58$&$<6.5\times 10^{-7}$\\
		$J/\psi\to D_s^-\mu^+\nu_\mu$&$19.3\pm2.0$&$18.4\pm0.8$&$31.8\pm4.8$&$35.4^{+5.0}_{-4.3}$&$17^{+7}_{-5}$&$37.5$&$55\sim57$&\\
		%\hline \hline
		\end{tabular}
	\end{ruledtabular}
\end{sidewaystable}		
	
\begin{table}[h!]
	
	\caption{Branching fraction of the nonleptonic $J/\psi\to D_{(s)}$ decays.}\label{Br2}
	\begin{ruledtabular}
		\begin{tabular}{@{\!\!}c@{}ccc@{\!}ccc@{}c@{}}	
			&\multicolumn{6}{c}{Theory }&Experiment\\
	\cline{2-7}
			\text{Decay}&\text{Unit}&\text{Our}&\text{BS}\cite{Wang2017}&\text{QCDSR}\cite{Wang2008}&\text{BSW}\cite{Dhir2013}&\text{QCDF}\cite{Sun2016}&\text{PDG}\cite{Navas2024}\\
			\hline
			$J/\psi\to D^+\pi^-+D^-\pi^+$&$10^{-11}$&$0.79\pm0.10$&$1.83^{+0.27}_{-0.25}$&$0.80\pm0.20$&$3$&$12.74^{+1.32}_{-0.80}$&$<7.0\times 10^{-8}$\\
			$J/\psi\to D^0\pi^0+\overline{D}^0\pi^0$&$10^{-12}$&$0.66\pm0.14$&$1.56^{+0.24}_{-0.21}$&&$2.4$&$7.00^{+4.12}_{-2.10}$&$<4.7\times 10^{-7}$\\
			$J/\psi\to D_s^+\pi^-+D_s^-\pi^+$&$10^{-10}$&$1.47\pm0.19$&$4.75^{+0.67}_{-0.59}$&$2.0^{+0.4}_{-0.2}$&$6.64$&$21.8^{+2.4}_{-1.6}$&$<1.3\times 10^{-4}$\\
			$J/\psi\to D^+K^-+D^-K^+$&$10^{-12}$&$0.58\pm0.07$&$1.31^{+0.19}_{-0.17}$&&$2.4$&$7.58^{+0.90}_{-0.58}$&\\
			$J/\psi\to D^0K^0+\overline{D}^0\overline{K}^0$&$10^{-11}$&$3.57\pm0.74$&$8.03^{+1.13}_{-1.03}$&$3.6^{+1.0}_{-0.8}$&$14.4$&$28.8^{+17.0}_{-8.8}$&$<1.7\times 10^{-4}$\\
			$J/\psi\to D_s^+K^-+D_s^-K^+$&$10^{-11}$&$1.08\pm0.14$&$3.12^{+0.42}_{-0.36}$&$1.6\pm0.2$&$4.8$&$12.36^{+1.40}_{-0.88}$&\\
			$J/\psi\to D^+\rho^-+D^-\rho^+$&$10^{-10}$&$0.65\pm0.11$&$1.13^{+0.16}_{-0.14}$&$0.42^{+0.18}_{-0.08}$&$1.44$&$4.24^{+0.54}_{-0.36}$&$<6.0\times 10^{-7}$\\
			$J/\psi\to D^0\rho^0+\overline{D}^0\rho^0$&$10^{-11}$&$0.55\pm0.12$&$0.960^{+0.135}_{-0.120}$&&$1.22$&$2.16^{+1.32}_{-0.70}$&$<5.2\times10^{-7}$\\
			$J/\psi\to D_s^+\rho^-+D_s^-\rho^+$&$10^{-9}$&$1.36\pm0.21$&$2.62^{+0.37}_{-0.32}$&$1.26^{+0.3}_{-0.1}$&$3.54$&$7.64^{+0.96}_{-0.64}$&$<1.3\times 10^{-6}$\\
			$J/\psi\to D^0K^{*0}+\overline{D}^0\overline{K}^{*0}$&$10^{-10}$&$2.92\pm0.58$&$4.75^{+0.68}_{-0.58}$&$1.54^{+0.68}_{-0.38}$&$5.02$&$8.18^{+5.10}_{-2.76}$&$<2.5\times 10^{-6}$\\
			$J/\psi\to D^+K^{*-}\!\!+D^-K^{*+}$&$10^{-12}$&$4.75\pm0.57$&$7.70^{+1.10}_{-0.93}$&&$8.4$&$22.8^{+3.6}_{-2.4}$&\\
			$J/\psi\to D_s^+K^{*-}\!\!+D_s^-K^{*+}$&$10^{-10}$&$1.01\pm0.13$&$1.67^{+0.24}_{-0.20}$&$0.82^{+0.22}_{-0.20}$&$1.94$&$4.00^{+0.60}_{-0.42}$&\\
			%\hline \hline
		\end{tabular}
	\end{ruledtabular}
\end{table}	
	
	In Table~\ref{Br2} we confront our results for the branching fractions of the nonleptonic $J/\psi$ decays with previous predictions BS \cite{Wang2017}, QCDSR  \cite{Wang2008}, BSW \cite{Dhir2013} and QCDF \cite{Sun2016}. We also give available experimental upper limits. We see that all theoretical predictions are about four orders of magnitude lower than the current experimental bounds \cite{Navas2024}. The largest branching fraction of order of $10^{-9}$ is predicted for the $J/\psi\to D_s^+\rho^-+D_s^-\rho^+$ decay.  In general our results  agree within error bars with QCDSR predictions \cite{Wang2008}, while BS  \cite{Wang2017} results are more than a factor of two higher. The QCDF \cite{Sun2016} gives the largest values.  These differences in the results with the BS \cite{Wang2017}, BSW \cite{Dhir2013} model and QCDF \cite{Sun2016} are explained by the magnitude of the form factors at $q^2=0$ given in  Tables~\ref{FFc1}--\ref{FFc2}. Note that the nonleptonic $J/\psi$ decays to the $D_{(s)}$ meson and a pseudoscalar meson $P$ are determined by the value of the single form factor $A_0(M_P^2)$ [see Eq.~(\ref{x1})], while all other form factors $V(M_V^2)$, $A_1(M_V^2)$ and $A_2(M_V^2)$ contribute to the branching fractions of the $J/\psi\to D_{(s)} V$ decays [see Eq.~(\ref{x2})].  
The nonleptonic decays involving neutral $\pi^0$ and $\rho^0$ mesons are additionally color (the coefficient $a_2$ instead of $a_1$) and isospin (the factor $1/\sqrt2$ in the neutral meson decay constant)  suppressed with respect to the corresponding charged channels.

{
In Tables V and VI we give the rough estimates of the uncertainties of our calculations. They include the $5\%$ uncertainty of the form factors, about 2\% uncertainties of the $J/\psi$ total decay width and of the CKM matrix elements $V_{cq}$ ($q=d,s$). The uncertainties of the nonleptonic decays also include $10\%$ and $20\%$ errors in the values of $a_1$ and $a_2$, respectively, as well as $1-2\%$ uncertainty in the decay constants $f_{P,V}$. This estimate does not include unknown uncertainties of the factorization and $N_c\to\infty$ limit. }

\section{Conclusion} \label{sec:con}

In the framework of the relativistic quark model based on the quasipotential approach the form factors of the semileptonic $J/\psi$ meson transitions into $D_{(s)}$ mesons were calculated. The relativistic effects including the wave function transformations from the rest to moving reference frame and contributions of the intermediate negative energy states were consistently taken into account. This allowed us to reliably calculate the form factors in the whole accessible kinematical range without additional approximations and extrapolations. The form factors were
expressed through the overlap integrals of the meson wave functions which are known from the previous meson mass spectrum calculations. The  transferred momentum, $q^2$, dependence of these form factors was explicitly determined in the whole accessible kinematic range.
These form factors were applied for the calculation of the differential and total decay rates of the $J/\psi$ semileptonic decays using the helicity formalism. For the nonleptonic $J/\psi$ decays the factorization approach in the limit of the number of colors $N_c\to \infty$ was employed. The calculated total
branching fractions are of order $10^{-9}\sim 10^{-12}$, which is well below the current experimental upper limits from PDG \cite{Navas2024} and recent data from the BESIII Collaboration \cite{Ablikim2021,Ablikim2022,Ablikim2024,Ablikim2024a,Ablikim2025,Ablikim2025a,Ablikim2025b}. 
However, the comparison of our results and other theoretical predictions with these data indicates that the increase of the statistics, which is expected from the future Super Tau Charm Factory \cite{Achasov2024,Barnyakov_2020,Peng:2020orp,Cheng:2022tog}, can result in the first measurement of the  $J/\psi$ weak decays. Such measurement will be important test of the existing theoretical approaches.
	
\begin{acknowledgments}
	The authors are grateful to  A.V. Berezhnoy for useful discussions. The work of Ivan S. Sukhanov was supported in part by the Foundation for the Advancement of Theoretical Physics and Mathematics ``BASIS'' grant number 23-2-2-11-1.
\end{acknowledgments}

\bibliography{Jpsi}

\end{document}